# Physical Conditions in the Orion HII Region

Gary J. Ferland
Physics, University of Kentucky
Lexington, KY 40506 USA
gary@pop.uky.edu

## Abstract

The Orion Nebula is the defining galactic HII Region, and has been well studied over a broad range of wavelengths. Many of its properties are well established. The gas-phase metallicity is roughly 2/3 of solar, close to that of the star cluster. There is no evidence that grains are destroyed anywhere in the Orion environment; refractory elements such as Ca and Fe remain heavily depleted. If the grains are well mixed with the gas then the Trapezium cluster's radiation field will drive the gas away, because of the grain opacity. This pushes the grain-bearing gas back against the photodissociation region (PDR). Radiation pressure is more important than winds in controlling the overall geometry. The nebula is in a state of energy equipartition between energy input from starlight and energy stored in gas pressure, kinetic outflow, and the magnetic field.

There is much that is not clear, however. How can gas flow away from the PDR into the face of a radiation field that produces such effective radiative acceleration? Grains are photoionized by the stellar radiation field, and this charge establishes a Coulomb drag that couples the grains (which feel the force of the radiation field) to the gas (which has a far lower opacity but most of the mass). The grains drift relative to the gas, with (model dependent) speeds of order a fraction of a km s$^{-1}$. The HII region has a physical thickness of roughly $10^{12}$ km. The grains can drift an appreciable fraction of the thickness of the ionized larger in a timescale of $\tau < 10^{13}$ s $\approx 10^5$ yr. The age of the current geometry is unknown since the motion of $\theta^1$ C Ori is unknown. The likely equilibrium is one where grains drift away from the illuminated face of the HII region and accumulate near the PDR. The ionized gas, having lost its opacity, then expands into the direction of the central star cluster, looses its emission measure, and eventually becomes invisible. The loss of photoelectric grain heating in parts of the HII region may cause a dramatic change in electron temperature, and provide a natural explanation for the "t$^2$" or temperature fluctuations phenomenon. The role of the magnetic field in controlling properties of the HII region is unclear, although a component of turbulence equal to the Alfvén speed has been seen. Clearly much work remains to be done.

# 1 Introduction and motivation

This paper summarizes recent efforts at understanding the Orion HII region. The Orion Nebula is a well-known naked-eye object, and has been well studied from the X-Ray through radio. It is the defining blister or champagne flow HII region, in which a star cluster irradiates the skin of a molecular cloud causing hot ionized gas to flow away (Zuckerman 1973; Balick et al. 1974). The geometry is the result of a mix of physical processes, including effects of the starlight, stellar winds, embedded grains, gas pressure, and tangled magnetic fields.

High-resolution images of Orion reveal spectacular detail (O'Dell & Wong 1996). Clearly understanding the rich fine-scales structure is beyond the most sophisticated of current numerical hydrodynamical simulations. However, on the largest scales the geometry is simple, and the physics prescribing it must be too. As an example, it is possible to reconstruct the geometry of the HII region using the surface brightness in emission lines. The most complete study is the extensive work of Wen & O'Dell (1995), and a figure from that paper is shown in Figure 1. The nebula is concave with the symmetry roughly centered on $\theta^1$ C Ori, the source of ionization. The lesson is that the overall structure of the nebula (its *climate*) appears to be both simple and governed by simplicity. However the fine structure, the arcs, jets, and filaments (the *weather*) are awesomely complex, and this must be the result of a mix of the very localized phenomena as well as the region's history (the *Butterfly in Tahiti* syndrome that makes weather prediction so difficult). This paper considers processes that determine the climate.

Orion is certainly worthy of study in its own right – it is a nearby HII region with active and vigorous star formation. However, there are two other reasons that make Orion a compelling textbook. First, it is a benchmark for the study of the chemical history of galaxies. The ISM is the most metal-rich of the galactic constituents, so Orion should tell us the current chemical state of local regions of the galaxy. In this way it serves as a case study for the extragalactic HII regions. These are vastly more complicated than Orion, ionized by a star cluster, spatially unresolved, and are the basis for understanding such things as galactic chemical abundance gradients and the primordial helium abundance.

A second reason is to understand the so-called Narrow Lined Region (NLR) of Active Galactic Nuclei (AGN). This is known to be a dusty photoionized gas with a density between $10^3 – 10^6$ cm$^{-3}$, (Osterbrock 1989), and occurs within the inner kpc of the host galaxy. Looking towards the center of our own galaxy we see large molecular clouds and associated HII regions on the same physical scale. Were we to place an active nucleus at the center of our galaxy we too would have a NLR – the surfaces of these molecular clouds. Orion is an opportunity to study the physics of this interaction first hand.

## 2 Some things we understand

Recent progress in understanding the geometry, thermal, ionization and chemical states, and grain properties of the Orion Nebula is described here. Some of this progress has resulted in new questions, which are summarized in the following section.

### 2.1 The geometry

It is possible to reconstruct the geometry of an emission-line region using the surface brightness in a recombination line. This is proportional to the product $n_e n_{ion} \alpha_{eff} h\nu L$ where the $n$'s are the electron and ion densities, $\alpha_{eff}$ is an effective recombination coefficient, and $L$ the path through the gas. This requires an independent measure of the density and temperature, but this is possible using ratios of forbidden lines. When Osterbrock and Flather (1959) originally performed the experiment they found that the path through the nebula was substantially smaller than its projected width. They assumed that the HII region was actually spherical, which required that the gas only partially filled a fraction ε of the path L. This so-called filling factor represented the hypothesized presence of unresolved mist of filaments and blobs, surrounded by vacuum.

We now know that the optical HII region is an illuminated atmosphere on the surface of a background molecular cloud (Zuckerman 1973; Balick et al. 1974), so we view a plane parallel geometry from a special perspective. The most careful reconstruction of the geometry is that done by Wen & O'Dell (1995), and one of their figures is shown in Figure 1. Most of the optical emission originates from an atmosphere flowing towards us from a molecular cloud that lies beyond the central stars. The geometry suggests overall simplicity. The HII region is convex with a center of symmetry below $\theta^1$ C Ori. The ionized gas is a thin high-pressure region behind the Trapezium, with the thickness flaring out as you move away. This is expected if the ionized layer is held to the face of the molecular cloud by stellar radiation pressure. Both radiation pressure and the surface brightness decrease as we observe regions more distant from the stars.

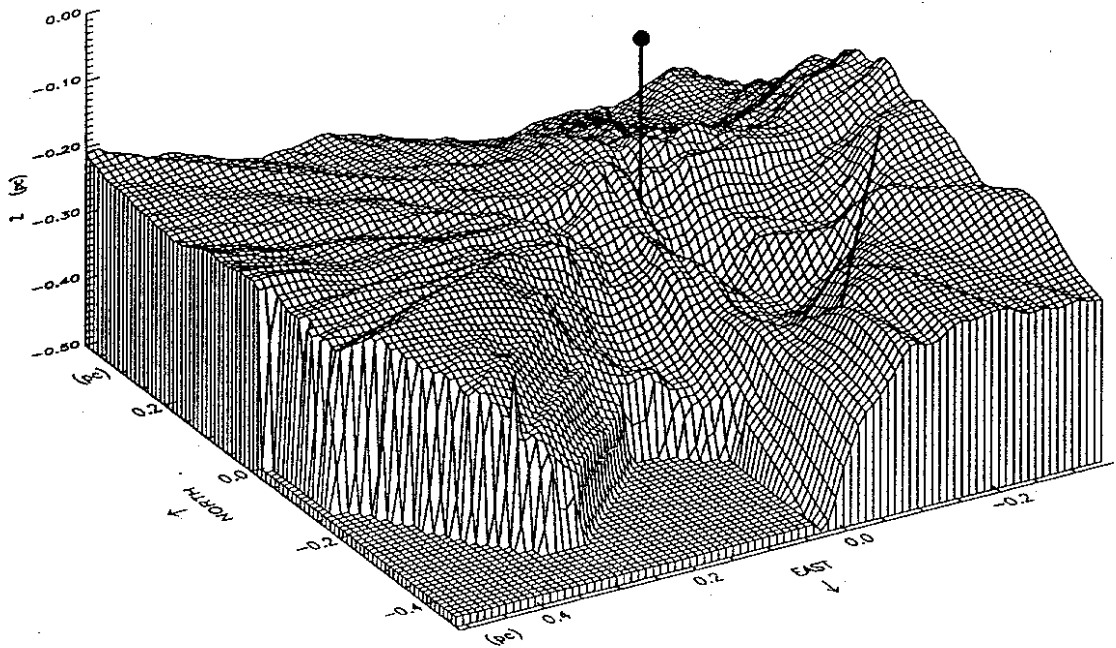

Figure 1 The reconstructed geometry of the main wall of the HII region. The z-axis is in parsecs, and θ¹ C Ori (the dot) is at a height of zero. We observe this geometry roughly face-on, from the +z direction. The HII region is an ionized layer on the surface of the molecular cloud that extends in the –z direction. From Wen & O'Dell 1995.

The surface brightness in a recombination line can also reveal the flux of ionizing photons that strike the cloud, and this is related to both the luminosity of the star in ionizing photons and its separation from the cloud. For a region close to the Trapezium Baldwin et al. (1991; hereafter BFM) measured an H$\beta$ surface brightness of $5.9 \times 10^{-12}$ erg cm$^{-2}$ s$^{-1}$ arcsec$^{-2}$. This corresponds to $4\pi J$ at the surface of the blister of 2.62 erg cm$^{-2}$ s$^{-1}$ or $6.4 \times 10^{11}$ H$\beta$ photons cm$^{-2}$ s$^{-1}$ if the layer emits isotropically. Actually the background molecular cloud partially reflects the layer's light, and this introduces a 10% correction. There are 8.6 hydrogen recombinations per H$\beta$ photon produced, so there are $5.0 \times 10^{12}$ recombinations per unit area and time. The nebula is radiation bounded (has a hydrogen ionization front) so there must be at least this many hydrogen ionizing photons striking the surface. Actually hydrogen must compete with grains in absorbing ionizing radiation, and the simulations presented by BFM suggest that only 63% of the ionizing radiation is absorbed by hydrogen. Applying this correction we find a flux of ionizing photons of $\Phi(H) = 0.8 \times 10^{13}$ cm$^{-2}$ s$^{-1}$, close to the number that came from BFM's numerical simulations, $1.0 \times 10^{13}$ cm$^{-2}$ s$^{-1}$. This number is one of the most robust that comes out of the photoionization simulations, and is important since, as we see below, it has major dynamical consequences.

The geometric reconstruction of Wen & O'Dell (1995) suggests that $\theta^1$ C Ori is 0.3 pc $\approx 9 \times 10^{17}$ cm away from the main layer of the HII region (see also O'Dell 1994). This distance and the flux quoted above suggest that the central star emits $Q(H) \approx 10^{50}$ ionizing photons per second. This is far higher than the standard estimate of $Q(H)$ for $\theta^1$ C Ori ($\approx 10^{49}$ s$^{-1}$; see O'Dell, 1998, and van der Werf & Goss 1989). This difference may be due to a combination of some uncertainty in the deduced distance between the cloud and star, and that fact that the original determinations of $Q(H)$ did not account for loss of ionizing photons onto grains. In the following I will take a separation of 0.2 pc, and discuss only results that are insensitive to this.

## 2.2 The origin of the geometry

The Orion HII Region is mainly ionized by the most massive star in the Trapezium cluster, $\theta^1$ C Ori. Its spectral class is roughly O6, corresponding to an effective temperature slightly below 40,000 K. Massive O stars tend to be windy, and $\theta^1$ C Ori has a mass loss rate of roughly $4\times10^{-7}$ $M_\odot$ yr$^{-1}$ at a velocity of $10^3$ km s$^{-1}$ (Howarth & Prinja 1989).

Photoionization analyses are sharply sensitive to $\Phi(H)$, the flux of hydrogen ionizing photons striking the illuminated face of the cloud. The models BFM computed suggested $\Phi(H) \approx 1.0\times10^{13}$ cm$^{-2}$ s$^{-1}$. This flux and the deduced continuum shape correspond to an energy flux in starlight of $1.0\times10^3$ erg cm$^{-2}$ s$^{-1}$. For comparison the kinetic energy associated with spherical mass loss from $\theta^1$ C Ori at a distance of 0.2 pc is $\rho v^2 / 2 \approx 3\times10^{-2}$ erg cm$^{-2}$ s$^{-1}$, so starlight is far more important than winds in this environment.

Much is known about the grain properties in the Orion region. A warm thermal infrared source is associated with the central stars. Refractory elements such as Ca and Fe are strongly depleted in the HII region (Baldwin et al. 1996). The grains in the layer in front of the Trapezium have a large ratio of total to selective extinction, indicating a large size distribution (Cardelli, Clayton, & Mathis 1989). The total grain cross section per proton is typically $10^{-21}$ cm$^2$. The radiative acceleration resulting from illumination by $\theta^1$ C Ori is $10^{-5}$ cm s$^{-2}$ for $\Phi(H) \approx 1.0\times10^{13}$ cm$^{-2}$ s$^{-1}$ (BFM). Starlight striking the grains carries enough momentum to push the ionized gas back against the molecular cloud. This starlight also photoionizes the grains which become an important additional source of gas heating. The Coulomb drag caused by the net positive charge of the grains couples them to the gas, although they drift relative to the gas with velocities of a fraction of a km/s.

BFM assumed hydrostatic equilibrium to compute the structure of the ionized layer. The density at each point was set by the balance between the integrated pressure of the absorbed starlight pushing gas away from the stars, with the local pressure determined by the gas density, electron temperature, and radiation pressure due to trapped lines. The pressure at the bottom of the ionized zone, near the ionization front, was equal to the radiation pressure from the stellar radiation. This geometry had simplicity as its primary motivation, and would be valid if the gas remains well coupled to the grains. Momentum from the starlight is actually absorbed predominantly by the grains, which then pulls the gas along with it. The grains drift relative to the gas, and this velocity is set by the balance between momentum absorbed from the starlight and the Coulomb drag between the grains and gas. The gas has most of the mass but little opacity. If all of this is correct then there is an important consequence – it is impossible for gas to flow away from the molecular cloud towards the Trapezium stars.

## 2.3 Orion's energy partition

The interstellar medium is in a state of general energy equipartition (Spitzer 1978), and if the Orion environment is old enough, it should attain this balance as well. Table 1 summarizes the energy partition of the region of the Orion Nebula near the Trapezium stars. The following components are present:

Table 1 Orion's Energy Partition

| Component | Energy density [K cm$^{-3}$] |
|---|---|
| Magnetic field | 5×10$^7$ |
| Alfvén waves | 5×10$^7$ |
| expansion | 6×10$^7$ |
| thermal energy | 9×10$^7$ |
| Total - nebula | 3×10$^8$ |
| Starlight | 3×10$^8$ |
| Winds | 2×10$^6$ |
| Total - stars | 3×10$^8$ |

*The magnetic field.* The magnetic field of the layer in front of the HII Region has been measured using Zeeman splitting of the 21 cm line, seen in absorption against the free-free continuum produced by the nebula (Troland et al. 1989; 1998). Peak values are typically 200 µG. This measures only the line of sight component of the field; the total field is twice this if the field is isotropic. The total energy density associated with a magnetic field is $B^2/8\pi$. This is not the magnetic field in the main HII region, but rather the "lid" discussed by O'Dell & Wen (1992).

Line broadening can also measure the magnetic field within the HII region itself. A magnetic field creates Alfvén waves with velocities given by $v_A^2 = B^2/4\pi\rho \approx 7.5\times10^{15}/n_e$ cm$^2$ s$^{-2}$ where $\rho$ is the density (gm cm$^{-3}$) and $v_A$ the Alfvén speed (Spitzer 1978). For the densest regions of the nebula the Alfvén speed is $\approx 9\,B_{400}\,n_4^{-1/2}$ km s$^{-1}$, where $B_{400}$ is the total magnetic field in units of 400 µG and $n_4$ the density in units of 10$^4$ cm$^{-3}$. Emission lines from the Orion HII region do exhibit an extra non-thermal component of broadening of just this size (Cansteñeda 1988; O'Dell 1998), suggesting that a similar magnetic field is present in the HII region itself. The energy associated with Alfvén waves with this density and field is 6×10$^{-9}$ erg cm$^{-3}$ ≈ 5×10$^7$ K cm$^{-3}$. Alfvén waves are non-dissipative, so although the energy associated with them is substantial, they do not contribute additional heating or cooling terms to the gas thermal balance.

*Expansion.* This represents the bulk expansion of the [OIII] emitting layer relative to the PDR. The energy density is $\rho v^2/2$. I took 10 km s$^{-1}$ and a density of 10$^4$ cm$^{-3}$.

*Thermal energy.* This is the heat content of the gas, *nkT*. A densty of 10$^4$ cm$^{-3}$ and a temperature of 9000 K were assumed.

*Starlight at HII region.* The flux of ionizing photons given above was assumed. Starlight is the major source of new energy going into the HII region.

*Winds.* Winds are often assumed to be a major energy source for gas near early type stars, but this does not appear to be the case for Orion. Howarth &

Prinja (1989) estimate the mass loss rate $\theta^1$ C Ori, as $4\times10^{-7}$ $M_\odot$ yr$^{-1}$ with a velocity of $10^3$ km s$^{-1}$. A cloud - star separation of $6\times10^{17}$ cm was assumed.

*Overall equilibrium.* The environment appears to be in rough energy equipartition, with the energy available in starlight balancing the energy reservoirs in the nebula. Stellar winds are relatively unimportant. The magnetic field *is* important, and so MHD effects may be too.

## 2.4 Some timescales

The Orion environment has obvious hydrodynamical complexity. But the nebula appears to have had time to reach energy equipartition, and the *overall* geometry appears simple (Figure 1). What is the steady state equilibrium of the Orion environment? What timescale would be needed to establish steady state, or the energy equipartition that appears to be present? Several timescales can be easily measured, but the most important, the time the star cluster has illuminated the blister from its current position, is poorly known.

*Thermal or cooling time.* This is nkT/$\Lambda \approx$ 150 years, where $\Lambda$ is the energy loss rate in the nebula and I used numbers from the BFM calculation.

*Expansion time.* This is the time required for the gas to move across the HII region. I used a layer thickness of $10^{17}$ cm and a flow speed of 10 km s$^{-1}$.

*Sound crossing time.* This is the time needed for the nebula to reach some sort of pressure balance. It is given by the ratio of the layer thickness to sound travel time. This is nearly equal to the expansion time since the ionized gas is observed to expand away from the PDR at roughly the sound speed.

*Grain drift time.* This is the time for the grains to drift across the ionized layer. I took a thickness of $10^{17}$ cm and a drift speed of 0.3 km s$^{-1}$.

*Recombination times.* This is the minimum time needed for the gas to reach photoionization equilibrium. It is roughly 1 / $n_e$ $\alpha_B \approx$ 150 years where $\alpha_B$ is the case B hydrogen recombination coefficient and I took a density of $10^4$ cm$^{-3}$.

*$H_2$ formation in the PDR.* I recomputed the photoionization model constructed by BFM, but extending the calculation until the point where nearly all carbon is in the form of CO. Ferland, Fabian, and Johnstone (1994) describe the methods. The code now checks many timescales and identifies the longest, since the calculation is only valid if the gas is in steady state. The slowest of the timescales is for $H_2$ formation, and is 10 million years. The $H_2$ part of the PDR chemistry is clearly not in equilibrium (Bertoldi and Draine 1996).

*Star-crossing time.* This is the most important timescale since the stars are the fundamental energy source for the nebula, and they control its structure. The question here is basically how long the Orion Nebula has existed in its present form. This is unknown, since the motion of $\theta^1$ C Ori is unknown (O'Dell 1998). Taking 2 km/s as a typical speed, the crossing would be ~$10^5$ years.

Table 2 Some timescales

| Property | $\tau$(years) |
|---|---|
| Thermal | 150 |
| Expansion | 3000 |
| Sound crossing | 3000 |
| Grain drift | 100,000 |
| Recombination | 150 |
| $H_2$ formation | $10^6$ |
| Star crossing | ??? |

The conclusion here is that the ionized part of the nebula should reach some sort of equilibrium in less than $10^4$ years. The star crossing time is unknown, but likely to be substantially longer than this. The PDR chemistry is clearly non-equilibrium, however.

## 2.5 Metallicity

A primary goal of the study of HII regions is to use them as probes of the chemical state of the ISM. Studies of the Orion Nebula agree that its gas-phase metallicity is roughly two thirds of solar (Peimbert and Torres-Peimbert 1977; BFM; Rubin et al. 1991; Osterbrock et al. 1992). This must be corrected for depletion onto grains, but the correction is not large. Recent work on abundances of stars in the Orion association agrees that the metallicity is below solar (Cunha, Smith, & Lambert 1995; 1998). Refractory elements are not condensed into grains in warm stellar atmospheres, of course.

The fact that the metallicity of the Orion environment is below solar is a challenge for simple models of galactic chemical enrichment (see the 1997 monograph by Pagel). The usual explanation is that the composition of the forming solar system was anomalously enriched by contamination from a passing supernova blast wave. There is also a growing consensus that ISM depletions have been overestimated by the assumption that the material begins with solar abundances (Snow & Witt 1996). This has further consequences for the composition and mean density of interstellar grains (Mathis 1996).

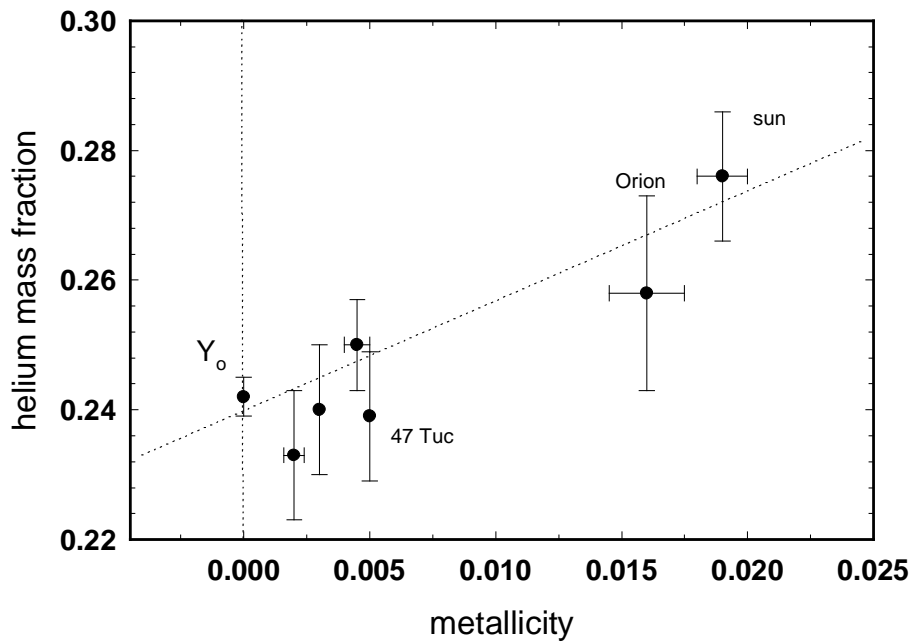

Figure 2 The helium mass fraction vs metallicity for a variety of objects, taken from BFM.

## 2.6 Grain survival

Do grains exist across all of the HII Region? It has long been known that grains do not exist close to the Trapezium star cluster (White, Schiffer, & Mathis 1980). Grains equilibrate at a temperature set by the balance between heating (mainly by starlight) and cooling (by radiation). Grains anywhere near the cluster would be too hot to produce the observed infrared continuum. This, together with observed scattering properties, led to the idea that grains had been destroyed within the HII region, since it was thought that gas filled the region surrounding the Trapezium.

BFM proposed that the HII region is pushed back to the wall of the blister by radiation pressure due to starlight pushing on the grains. In this picture no grains are close to the stars because no gas is either. Their predicted infrared continuum was in good agreement with much of the infrared observations. The luminosity of the thermal infrared emission requires that grains do see the full luminosity of the Trapezium cluster. All of this suggests that the Orion environment is dusty, with no particular suggestion that any regions are cleansed of grains.

Gas phase abundances offer a second clue to grain survival. Refractory elements such as iron and calcium are depleted by several orders of magnitude in the ISM (Savage & Sembach 1996). Were gas-phase calcium present in the HII region with a solar abundance then the [CaII] $\lambda\lambda 7291, 7324$ lines would be among the strongest in the spectrum (Ferland 1993, Kingdon, Ferland, & Feibelman 1995). They are not seen, showing that the Ca depletion is very high. The observed [FeII] emission is consistent with an iron depletion of an order of magnitude (Baldwin et al. 1996). There is no evidence of selective depletion in high ionization stages of ionization of silicon. While this suggests that grains are not destroyed anywhere, it does not tell us where they may be located.

## 2.7 A ray through the Orion blister

Figure 3 shows the ionization and thermal structure of the Orion HII region and PDR. This was computed using the parameters described by BFM and Baldwin et al. (1996), but extends through the PDR to the point where carbon has gone totally to CO. The model has constant total pressure, and is a hydrostatic layer that results from a balance between the force due to radiation pressure from the Trapezium stars, and back pressure due to radiation pressure of trapped lines and thermal gas pressure. The $H^+$ zone is nearly isothermal, as is typical of photoionization calculations. Because the model is assumed to maintain constant pressure, the density in the PDR is high, roughly $10^6$ cm$^{-3}$.

Figure 4 shows the spectrum emitted by this structure. The optical through IR spectrum clearly has a wealth of information. The precise form of the infrared thermal continuum is the primary indicator for the location of the warm grains.

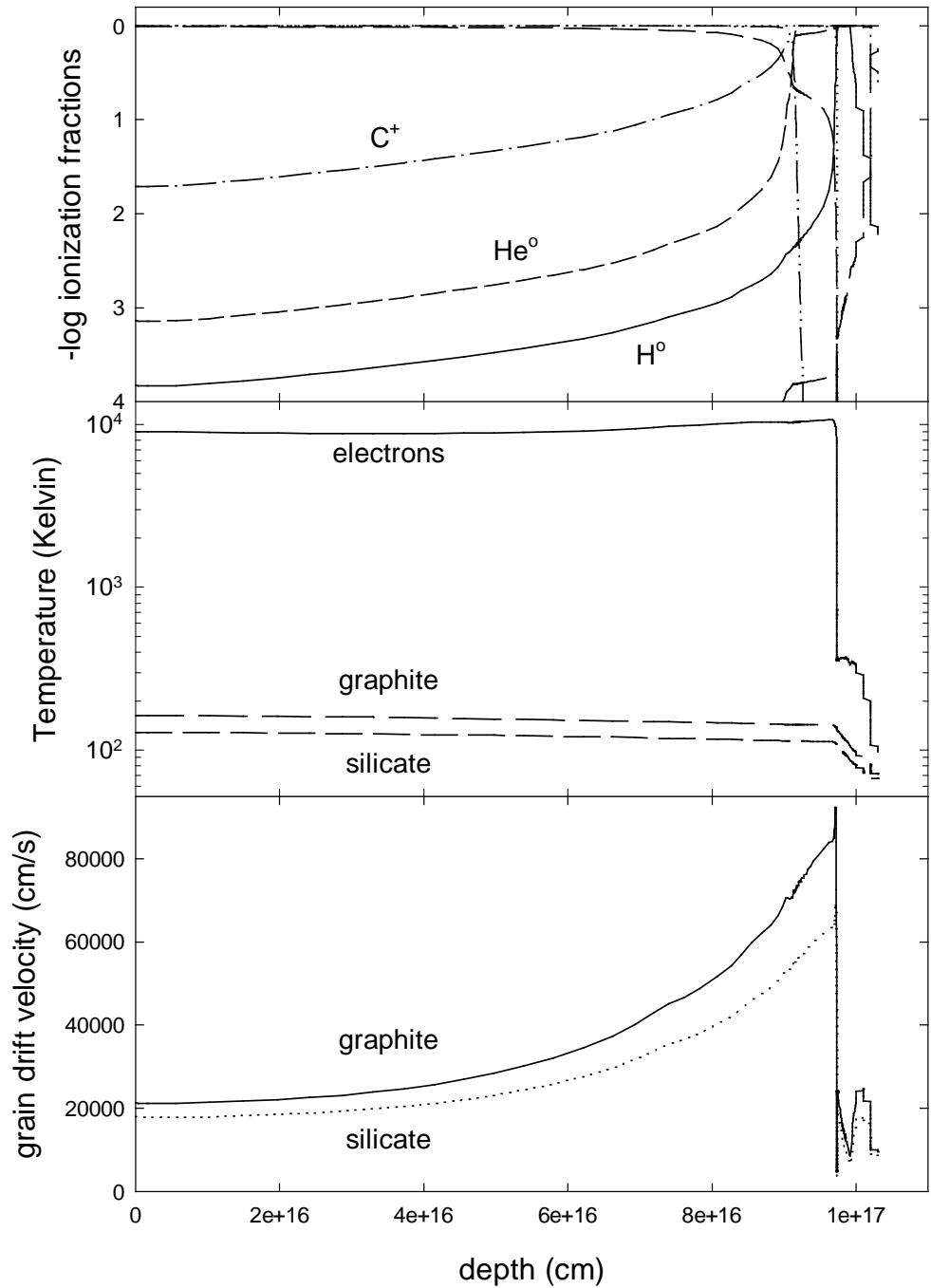

Figure 3 The ionization and thermal structure of the Orion HII region and PDR. The grain temperatures and drift velocities are shown in the lower two panels.

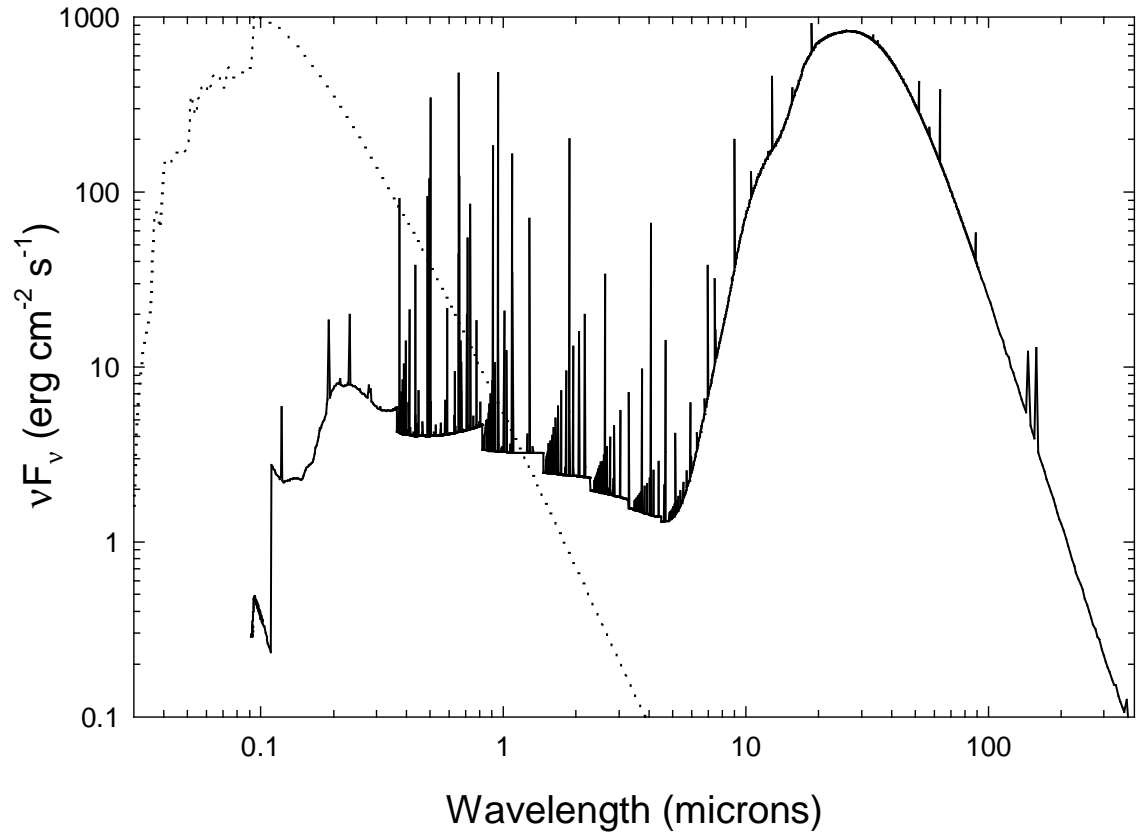

Figure 4 The spectrum emitted by the structure shown in the previous figure. The incident stellar continuum is the dotted line, and the predicted nebular emission is shown as a solid line.

For instance, silicates are warm enough for the 10 micron feature to be weakly present. High spatial resolution maps in the 4 – 40 micron region will tell us much about the equilibrium state and position of the grains.

# 3 Some things we don't understand

Orion is bright and has a large apparent diameter. As a result its properties can be determined with a clarity that is impossible in most objects. Careful scrutiny has revealed many open problems. Some are discussed here.

## 3.1 The $t^2$ dilemma

For many years the Orion HII region was visualized as a mist of unresolved blobs, with the ensemble characterized by a "filling factor". The origin of these blobs was unknown. It also seemed physical that these components might not have the same temperature. Peimbert (1967) introduced the concept of "temperature fluctuations", small-scale changes in the electron temperature over regions that might otherwise be expected to be isothermal. These fluctuations are characterized by $t^2 \equiv (\Delta T/T)^2$, the square of the dispersion of the electron temperature about its mean.

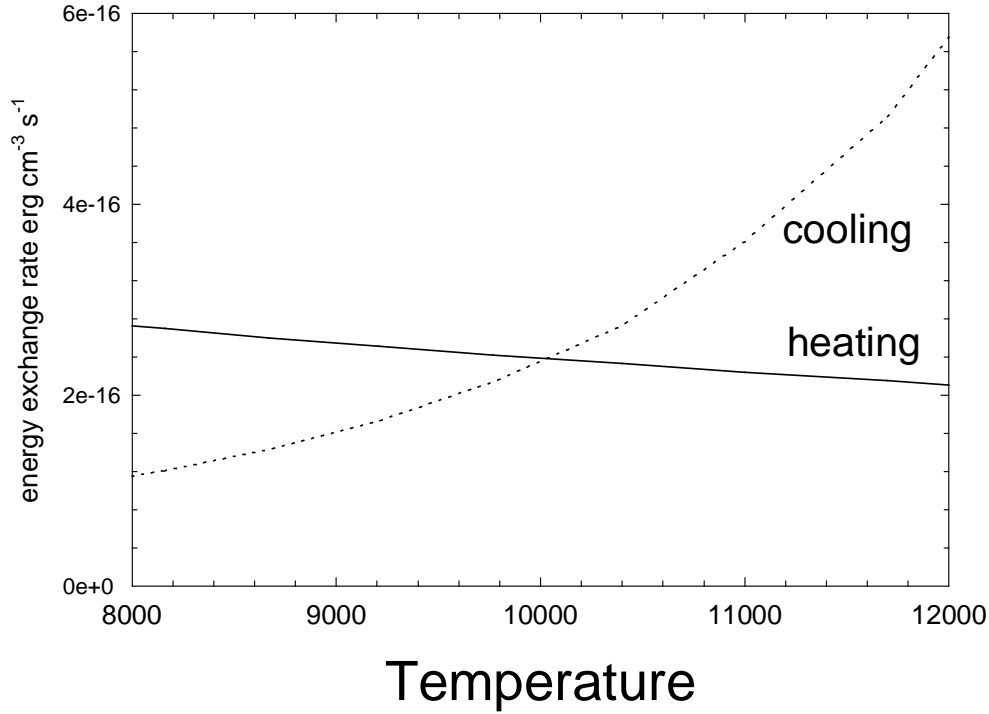

Figure 5 The photoelectric heating and total cooling for a region within the [OIII] zone of the Orion Nebula.

Theoretically, $t^2$ is controversial since a gas in photoionization equilibrium is nearly isothermal (see Figure 3 and also Kingdon & Ferland 1995). This is due to the steep form of the gas cooling function (see Figure 5). The heating rate is proportional to $n^2 \alpha \langle h\nu - h\nu_o \rangle$ and decreases slowly with increasing temperature because the recombination coefficient does. The cooling rate increases exponentially since it is predominantly due to collisionally excitation of emission lines. Density fluctuations cannot produce $t^2$ either. The local heating and cooling rates have the same density squared dependence. The heating – cooling balance is not affected by density changes unless the ionization of the heavy elements changes significantly or the cooling lines become collisionally suppressed (Viegas & Clegg 1994).

Although $t^2$ is not theoretically expected, there is observational evidence, based on pairs of different temperature indicators, that it does exist, and that it can be as large as 0.03 in Orion (Rubin et al. 1998). If this is correct then it has profound implications for the equilibrium of a photoionized gas. Figure 5 shows the heating and cooling rates for a region midway within the [OIII] region of the Orion model shown above. The equilibrium electron temperature is the result of the balance between these two, and is about $10^4$ K for this spot. A $t^2$ of 0.03 requires that dT/T be 0.17, or that the heating-cooling balance fluctuates by roughly 1500 K from equilibrium.

Figure 5 shows that the heating - cooling mismatch 1500 K away from equilibrium is roughly a factor of two. This requires that extra heating terms comparable to the photoionization rate come and go over small scales, and that it happen more quickly that the 150 year cooling time of the nebula. This seems to require a conspiracy since a totally independent heating source should be orders of magnitude larger or smaller than the photoionization rate, and not follow photoionization consistently within a factor of two. It is hard to think of a mechanism that would both do this, and replenish the fluctuations faster than the cooling time.

Although it is hard to produce temperature fluctuations theoretically, if $t^2$ is real then it may solve the well-known dilemma that abundances determined from recombination lines are higher than those determined from collisionally excited lines (Liu et al. 1995). In a gas with significant $t^2$, those lines with weak temperature dependencies (such as recombination lines) are affected the least, while those with sharp dependencies (collisionally excited lines) will be heavily weighted towards the hottest gas. The effect is that the hydrogen recombination spectrum will indicate a lower temperature than the collisionally excited lines, and that collisionally excited lines indicate a lower abundance than heavy element recombination lines. This is commonly observed, adding to the substantial body of evidence that something has gone wrong.

Nobody has produced a hypothesis to explain the origin of the $t^2$ phenomenon, or why the collisional and recombination abundances do not agree. There may be an essential piece of physics that has been left out of our understanding of nebulae.

## 3.2 Flow geometry

An O star is only slightly below the Eddington limit even for electron scattering opacity alone. If the HII region is dusty then its opacity is increased by many orders of magnitude and the gas will be forced away from the stars. The radiative acceleration predicted by BFM was large, $\approx 10^{-5}$ cm s$^{-2}$. If this were the whole story then the HII region gas would form a static atmosphere in hydrostatic equilibrium on the wall of the blister.

This cannot be the whole story, however. Lines coming from the PDR have radial velocities close to that of OMC1, presumed to be the center of mass of the molecular cloud. There is a clear radial velocity gradient increasing with higher ionization species, ie., [OI] to [OII] to [OIII] (Balick et al. 1974, Zuckerman 1973). The gas appears to be simply freely expanding away from the PDR as it is heated and ionized by the ionizing radiation field of $\theta^1$ C Ori. This is impossible if grains are well mixed within the gas. Clearly something is wrong – grains mixed with gas cannot flow towards the stars. This suggests that the grain geometry may be more complicated than previously thought, as described next.

## 3.3 Grain drift velocity, and Orion's age

Figure 3 shows the drift velocity for the two classical grains species included in the BFM calculation. Drift velocities are roughly several tenths of a km/s near the illuminated face of the cloud, where the grains are highly ionized and so well-coupled to the gas. Deeper in the cloud the grains are shielded and more neutral, and so drift more quickly. Taking 0.3 km s$^{-1}$ as a typical drift velocity and $10^{12}$ km as a thickness, the timescale for the grains to drift across the ionized layer is $\approx 10^5$ years. If the nebula has an age comparable to this then the dust and gas will not be well mixed.

How old is the current geometry? The motion of the stars relative to the gas sets this. O'Dell (1998) discusses the major uncertainties concerning the motion of $\theta^1$ C Ori. No reliable conclusions can be reached, but an age in excess of $10^5$ years seems likely if the motion of others stars in the environment is used.

Depending on the age, grains may have drifted past some parts of the HII region. These grains will then build up in deeper regions of the cloud, near or within the PDR. The intermediate situation could have the part of the layer nearest the illuminated face cleansed of grains, with deeper regions having a substantially larger than expected dust to gas ratio. This could have dramatic effects on the thermal structure of the cloud, since the primary effects of the extra grains would be to provide an additional component of photoelectric heating. The effect could be to increase the photoelectric heating at the factor of two level, over scales small compared to that of the [OIII] region. This is precisely what is needed to produce a significant t$^2$ (see Figure 5 and associated discussion).

What is the time-steady solution? Even this is unclear. There will be a flow of ionized gas past the dusty region, that will be set by the grain drift velocity, and constitutes a mass flux of $\approx m_p n \mathrm{v}_{drift}$ gm cm$^{-2}$ s$^{-1}$ $\approx 10^{-15.5}$ gm cm$^{-2}$ s$^{-1}$. Once gas has moved clear of the grains, it will expand freely into the central cavity. This must be the origin of the gas that is blue-shifted relative to the PDR. The visibility of an emission-line region is proportional to its emission measure, $E_m = n^2 V$. In the simplest case of an infinite plane-parallel wall with no further acceleration, the gas densty will not decrease further and so will retain its emission measure. If the gas density falls, either because the gas accelerates or expands, then it could eventually become undetectable, at least as an emission component. This appears to be necessary to explain the observations.

The situation is clearly a rich environment, and we now have only a hint of the properties of the true time-steady solution. Complete numerical simulations will require a combination of the entire plasma and grain physics, as well as the magneto-hydrodynamics. This is well beyond the capabilities of today's codes.

Even were it possible to perform the calculation, the simulations would be no better than our knowledge of the motion of θ$^1$ C Ori. This is the single biggest question, basically the question the Orion Nebula's age.

## 4 The future

Much work remains to be done. Several lines of investigation are currently underway. These include:

- More realistic grain calculations, including grain size distribution, are now possible. The simulations in BFM were at the limit of what could be done in 1990, but today's computers are much faster. Work is now underway to do size-specific grain distributions. It seems likely that the grains will sort themselves out by size across the HII region, and that grains will build up near the hydrogen ionization front.

- PAHs have been mapped across the Orion bar, and disappear when the gas becomes ionized. Is this related to the grain drift? Have they been destroyed or simply pushed back? Do other lines of sight through the Orion Nebula force similar conclusions about the location of the PAHs?

- What is the role of the magnetic field? It is in rough energy equipartition with other energy sources, and the Alfvén velocity accounts for the extra component of turbulence seen in the emission lines. This overall energy equipartition is the result of energy flowing back and forth between various constituents. There must be some regions where the magnetic field is more intense than average and so controls the matter. Could this produce some of the fine scale structure observed in the HII region? If the extra component of turbulence is the Alfvén speed then the width of emission lines from heavy ions such as [SIII] can be used to measure $B$ in the HII region. These magnetic field maps could be done on the sub arcsecond scale with HST. Does the magnetic field scale with the any of the observed structures?

- The physical conditions in the lid. This is one of the very few regions where we have spatial maps of the magnetic field (Troland et al. 1989; 1998). The Orion region seems to have a rough equipartition between magnetic and other energy forms. What are the conditions in the lid, and how do they change with the magnetic field? Does the density, temperature, or turbulence, scale with observed magnetic field? This could be determined by UV absorption line studies of stars behind the lid.

- Maps in the 4 -40 micron region. Photoionization calculations reproducing the optical/IR/UV emission line intensities make explicit predictions of the thermal infrared emission. The only published maps are decades old and with very poor spatial resolution. Accurate maps could help identify the location of the emitting grains, and determine their physical properties.

- The motion of $\theta^1$ C Ori. This is the single biggest piece of the puzzle. O'Dell (1998) summarizes the bizarre and contradictory evidences concerning its motion. Intricate MHD simulations of the full environment will only be meaningful once the motion of this star is known.

I want to thank Bob O'Dell for his careful reading and helpful comments on this paper. My colleague Tom Troland educated me on the importance of magnetic fields. Peter Martin taught me everything I know about grains. Research in Nebular Astrophysics at the University of Kentucky is supported by the NSF through grant AST 96-17083.